\documentclass[a4paper,12pt]{article}
\usepackage[margin=1in]{geometry}
\usepackage{color}
\definecolor{indigo}{RGB}{0,0,120}
\usepackage[colorlinks=true, linkcolor=indigo, citecolor=blue, urlcolor=indigo]{hyperref}
\usepackage{amsmath}
\usepackage{amsfonts}
\usepackage{txfonts}

\newcommand{\beq}{\begin{equation}}
\newcommand{\eeq}{\end{equation}}
\newcommand{\beqs}{\begin{eqnarray}}
\newcommand{\eeqs}{\end{eqnarray}}
\newcommand{\half}{\frac{1}{2}}
\newcommand{\dd}[2]{\frac {\partial #1}{\partial #2}}
\newcommand{\pdr}{\partial}
\newcommand{\bra}{\langle}
\newcommand{\ket}{\rangle}
\newcommand{\grad}{\nabla}

		\def\g{\gamma}
\def\del{\delta}	\def\D{\Delta}
\def\la{\lambda} 	\def\tht{\theta}

\usepackage{titlesec}
\titleformat{\section}{\normalsize\bfseries}{\thesection}{1em}{}
\titleformat{\subsection}{\small\bfseries}{\thesubsection}{1em}{}
\titleformat{\subsubsection}{\small\bfseries}{\thesubsubsection}{1em}{}

\newcommand{\bfG}{{\bf G}}
\newcommand{\bfF}{{\bf F}}
\newcommand{\bfv}{{\bf v}}

\newcommand{\bfU}{{\bf U}}
\newcommand{\bfr}{{\bf r}}
\newcommand{\bfa}{{\bf a}}
\newcommand{\bfn}{{\bf n}}

\begin{document}

%-------------------------------------

\title{\LARGE Higgs Mechanism and the Added-Mass Effect}
\author{{\sc Govind S. Krishnaswami and Sachin S. Phatak}\\ \\
{Chennai Mathematical Institute,}\\ {SIPCOT IT Park, Siruseri 603103, India.}\\ 
{Email: {\tt govind@cmi.ac.in, phatak@cmi.ac.in}}}

\date{February 2, 2015}

\maketitle

\centerline{Published in {\em Proc. R. Soc. A} {\bf 471}: 20140803 (2015); E-print arXiv:1407.2689}

\begin{abstract}

In the Higgs mechanism, mediators of the weak force acquire masses by interacting with the Higgs condensate, leading to a vector boson mass matrix. On the other hand, a rigid body accelerated through an inviscid, incompressible and irrotational fluid feels an opposing force linearly related to its acceleration, via an added-mass tensor. We uncover a striking physical analogy between the two effects and propose a dictionary relating them. The correspondence turns the gauge Lie algebra into the space of directions in which the body can move, encodes the pattern of gauge symmetry breaking in the shape of an associated body and relates symmetries of the body to those of the scalar vacuum manifold. The new viewpoint is illustrated with numerous examples, and raises interesting questions, notably on the fluid analogs of the broken symmetry and Higgs particle, and the field-theoretic analogue of the added mass of a composite body.

\end{abstract}


%============================================

\normalsize

%-----------------------
\section{Introduction}
\label{s-intro}
%-----------------------

In the recently confirmed \cite{ATLAS,CMS} Higgs mechanism \cite{Anderson,EnglertBrout,Higgs,GuralnikHagenKibble}, the otherwise massless carriers of the weak force ($W^\pm$, $Z$ gauge bosons) acquire masses by interacting with the Higgs medium. It is tempting to look for analogies where a body gains mass while moving through a fluid, to complement standard examples of (Abelian) mass generation for photons in a superconductor or plasma. Fluid analogies are often unsatisfactory, since they suggest resistive effects which are not present in the Higgs mechanism. However, McClements and Thyagaraja\cite{Thyagaraja} recently pointed out that a dissipationless fluid analog for the Higgs mechanism is provided by the added-mass effect. In its essence, this effect goes at least as far back as Green and Stokes (see Art.~92 in \cite{Lamb}). To impart an acceleration $\bfa = \dot \bfU(t)$ to a body of mass $m$ immersed in an inviscid, incompressible and irrotational fluid, one must apply a force exceeding $m {\bf a}$, since energy must also be pumped into the induced fluid flow. The added force $F^{\rm add}_i =  \mu_{ij} a_j(t)$ is proportional to the acceleration, but could point in a different direction, as determined by the added-mass tensor $\mu_{ij}$. $\mu_{ij}$ depends on the fluid and shape of the body, but not on its mass distribution, unlike its inertia tensor. For example, the added-mass tensor of a sphere is $\del_{ij}$ times half the mass of displaced fluid. So an air bubble accelerated in water `weighs' about $\frac{\rho_{\rm water}}{2 \rho_{\rm air}} \approx 400$ times its actual mass. The added-mass effect is different from buoyancy: when the bubble is accelerated horizontally, it feels a horizontal opposing {\em acceleration reaction force} $\bfG = - \bfF^{\rm add}$ aside from an upward buoyant force which is independent of $\bfa$ and equal to the weight of fluid displaced.

Here, we develop a novel and precise physical analogy between the added-mass and Higgs mechanisms. It is not a mathematical duality like the high temperature-low temperature Kramers-Wannier duality in the Ising model or the AdS/CFT gauge-string duality, but provides a fascinatingly new viewpoint on fluid-mechanical and gauge-theoretic phenomena. We discover a way of associating a rigid body to a pattern of spontaneous symmetry breaking (SSB). We call this the {\it Higgs Added-Mass (HAM) correspondence}, it applies both to Abelian and non-Abelian gauge models. Consider a $3+1$ dimensional Yang-Mills theory with $d$-dimensional gauge group $G$, which spontaneously breaks to a subgroup $H$ when coupled to scalars $\phi$ in a specified representation of $G$, subject to a given $G$-invariant potential $V$. The correspondence relates this to a rigid body accelerated (for simplicity) through a non-relativistic, inviscid, incompressible (constant density) irrotational fluid which is asymptotically at rest in $\mathbb{R}^d$. The Lie algebra $\underline{G}$ plays the role of the space through which fluid flows (with location of body as origin). In particular, the ($3+1$) space-time dimension of the gauge theory is unrelated to $d$. The fluid is the analog of the scalar field, while the rigid body plays the role of the vector bosons. Moreover, we propose a fluid analog for the Higgs particle. The correspondence proceeds through the respective mass matrices, and relates symmetries on either side, as exemplified by numerous examples that we present. 

We begin this paper in \S \ref{s-added-mass-effect} with a description of the added-mass effect, followed in \S \ref{ssb-patters-rigid-bodies} with a brief statement of the correspondence and several examples of SSB patterns and their corresponding rigid bodies. In each case, rotation and reflection symmetries of the body are related to symmetries of $G/H$, endowed with a metric implied by the vector boson mass matrix. Based on these examples, we present in \S \ref{s-Dictionary}, a detailed dictionary relating various quantities/phenomena on either side of the correspondence. The reader interested in a summary of the correspondence may start with \S \ref{s-Dictionary}. We conclude in \S \ref{Discussion} with a discussion of interesting questions that the new viewpoint raises.

% \vspace{-.5cm}

%------------------
\section{The Added-Mass Effect}
\label{s-added-mass-effect}
%------------------

% \vspace{-.4cm}

Perhaps the simplest example of the added-mass effect is in 1-dimensional flow. Consider an arc-shaped rigid body of length $L$ surrounded by ideal fluid filling the circumference of a circle of radius $R$. Incompressibility $\pdr_\tht v(\tht, t) = 0$, along with impenetrability of the body imply that the flow velocity $v$ everywhere is the same as that of the body $v = U(t)$. The rate of increase in flow kinetic energy
\begin{equation}
 \frac{d}{dt} \int_{\rm fluid} \half \rho \: v^2 \:  R d \tht = \rho (2\pi R - L) U \dot U
\end{equation}
must equal the power supplied by the added force $F_{\rm add} U(t)$. Thus, $F_{\rm add}$ is proportional to the acceleration of the body, which gains an added-mass $\mu = \rho (2\pi R - L)$. $\mu$ being equal to the total mass of fluid is peculiar to this one dimensional toy-model; this is why we choose a circular flow domain instead of the whole real line.

More generally, following Batchelor \cite{Batchelor}, consider incompressible (constant density) 3d {\it potential flow} around a simply-connected rigid body executing {\it purely translational} motion at velocity ${\bf U}(t)$ (see \S \ref{app-added-mass-compress-potn-flow} for extension to compressible flows). We restrict to the case where external forces do not cause the body to rotate. The fluid, assumed asymptotically at rest, has velocity ${\bf v} = \grad \phi$ with $\grad \cdot \bfv = \grad^2 \phi = 0$. $\phi$ is determined by impenetrability: $\grad \phi \cdot {\bf n} = {\bf U}(t) \cdot { \bf n}$ where $\bf n$ is the unit outward normal on the body's surface $A$. The boundary condition constrains $\phi$ to be linear in $\bf U$, which allows us to write $\phi({\bf r}) = {\bf U} \cdot {\bf \Phi}({\bf r})$ where we call ${\bf \Phi}({\bf r})$ the potential vector field. $\bf \Phi(\bfr)$ depends on the shape of the body, but not on its velocity $\bfU$. As time progresses, ${\bf \Phi} = {\bf \Phi}({\bf r} - {\bf r_0}(t))$ where ${\bf r_0}(t)$ is a convenient reference point in the body. For a sphere of radius $a$ instantaneously centered at the origin, 
	\beq
	\phi({\bf r}) = - \frac{a^3}{2 r^3} \bfU(t) \cdot \bfr 
	\quad \text{and} \quad
	{\bf \Phi} = - \frac{a^3}{2} \frac{\hat r}{r^2}.
	\eeq
Bernoulli's equation 
	\beq
	p + \half \rho v^2 + \rho \dd{\phi}{t} \: = \: {\rm constant}(t),
	\label{e-bernoulli-eqn}
	\eeq
allows us to express the total pressure force on the body as 
	\beq
	{\bf F} \; = \; - \int_{A} p \, {\bf n} \, dA = \rho \int_{A} \left( \frac{\partial \phi }{\partial t} + \frac{1}{2} v^2 \right) {\bf n} \, dA.
	\eeq
Using the factorization $\phi = {\bf U} \cdot {\bf \Phi}$, we write $\bfF$ as a sum of an acceleration reaction force $\bf G$ and an acceleration-independent ${\bf G}'$
	\beq
 	{\bf F} = \rho \int_{A} {\bf\dot{U}} \cdot {\bf \Phi} \: {\bf n} \: dA
	+ \int_{A} \left[ \half \rho v^2 - \rho {\bf U} \cdot {\bf v} \right] \: {\bf n} \: dA \equiv \bfG + \bfG'.\label{e-G-G-prime}
	\eeq
$\bfG'$ vanishes in fluids asymptotically at rest in $\mathbb{R}^3$ \cite{Batchelor}. Using a multipole expansion for $\phi$ (see appendix \ref{app-d-dim-addedmass}), one estimates that $\bfG'$ can be at most of order $1/R$ in a large container of size $R$. It is as if fluid can hit the container and return to push the body. We ignore this boundary effect. When acceleration due to gravity $\bf g$ is included, ${\bf G}'$ features a buoyant term $- \rho V_{\rm body} \, {\bf g}$ equal to the weight of fluid displaced, which we suppress. Thus, the acceleration reaction force is
\begin{equation}
 G_i = - \mu_{ij} \dot U_j \quad \textmd{where}\quad 
 \mu_{ij} = - \rho \int_{A} \Phi_j \: n_i \: dA.
\end{equation}
The added-mass tensor $\mu_{ij}$ is a direction-weighted average of the potential vector field $\bf \Phi$ over the body surface. It is proportional to the fluid density and depends on the shape of the body surface. $\mu_{ij}$ may be shown to be time-independent and symmetric. The rate at which energy is pumped into the fluid is $- {\bf G} \cdot {\bf U}(t) = \mu_{ij} \dot U_j U_i(t)$. Thus the flow kinetic energy may be expressed entirely in terms of the body's velocity and added-mass tensor (It follows that the added-mass tensor $\mu_{ij}$ is a positive matrix):
    \beq
    \half \int_V \rho v^2 \: dV \; = \; \frac{1}{2}\mu_{ij} U_i U_j.
    \eeq

To a particle physicist, mass generation in a medium sounds like the Higgs mechanism, and an added-mass tensor is reminiscent of a mass matrix. To uncover a precise correspondence between these phenomena, it helps to have explicit examples. By solving potential flow around rigid bodies, one obtains their added-mass tensors. We will relate these rigid bodies and their added-mass tensors to specific patterns of spontaneous gauge symmetry breaking. For a $2$-sphere of radius $a$, $\mu_{ij} = \frac{2}{3} \pi a^3 \rho \del_{ij}$ is isotropic. The added-mass of a sphere is half the mass of fluid displaced, irrespective of the direction of acceleration. For an ellipsoid $\frac{x^2}{a^2} + \frac{y^2}{b^2} + \frac{z^2}{c^2} =  1$, $\mu_{ij}$ is diagonal in the principal axis basis. If $a > b > c$, then the eigenvalues satisfy $\mu_x < \mu_y < \mu_z$. Roughly, added-mass grows with cross-sectional area presented by the accelerating body. In its principal axis basis \cite{Lamb}
	\beq
	(\mu_x, \mu_y, \mu_z) = \frac{4}{3}\pi\, abc\, \rho \, \left(\frac{\alpha}{2-\alpha}, \frac{\beta}{2-\beta}, \frac{\gamma}{2-\gamma} \right),
	\eeq
where 
    \beq
    \alpha = abc\, \int_{0}^{\infty} (a^2 + \lambda)^{-1} \D^{-1} d\lambda \quad
    \text{with} \quad
    \Delta = \sqrt{(a^2 + \lambda)(b^2 + \lambda)(c^2 + \lambda)}
    \eeq
and cyclic permutations thereof. In particular, for an ellipsoid of revolution with $a = b$, the corresponding pair of added-mass eigenvalues coincide $\mu_x = \mu_y$. On the other hand, by taking $c \to 0$ we get an elliptic disk, for which two added-mass eigenvalues $\mu_x$ and $\mu_y$ vanish. These correspond to acceleration along its plane. With impenetrable boundary conditions, an elliptic disk does not displace fluid or feel an added-mass when accelerated along its plane. The third eigenvalue $\mu_z$, for acceleration perpendicular to its plane, is $\frac{4}{3}\pi \rho ab^2 E( 1 - b^2/a^2 )^{-1}$, where $E(m)$ is the complete elliptic integral of the second kind. Taking $a=b$, the principal added-masses of a circular disk are $(0,0,(8/3)\rho a^3)$. Shrinking the elliptical disk further, a thin rod of length $2a$ has no added-mass. Irrespective of which way it is moved, it does not displace fluid with impenetrable boundary conditions. The same is true of a point mass or any body whose dimension is less than that of the flow domain by at least two (codimension $\geq 2$). For an infinite right circular cylinder, the added-mass per unit length for acceleration perpendicular to its axis is equal to the mass of fluid displaced. If the axis of the cylinder is along $z$, then the added-mass tensor per unit length is $\frac{\mu_{ij}}{L} = \pi a^2 \rho\; {\rm diag}(1,1,0)$, where $a$ is its radius. Though these examples pertain to three dimensional flows, the added mass effect generalizes to rigid bodies accelerated through plane flows as well as flows in $4$ and higher dimensions. The case of plane flow is well-known and treated for instance in \cite{Batchelor}. For example, an elliptical disk with semi-axes $a, b$ accelerated through planar potential flow has an added mass tensor $\mu_{ij} = \pi \sigma a b \del_{ij}$ where $\sigma$ is the (constant) mass of fluid per unit area. In appendix \ref{app-d-dim-addedmass} we develop the formalism for the added mass effect in $d \geq 3$ dimensions. This will be used in the following sections where we relate the added mass effect in $d$-dimensional flows to spontaneous breaking of a $d$-dimensional gauge group $G$.

% \vspace{-.5cm}

%=========================================================
\section{SSB Patterns and their Rigid Bodies} 
\label{ssb-patters-rigid-bodies}
%=========================================================

% \vspace{-.4cm}

In the simplest version of the Higgs mechanism, a $G = $U$(1)$ gauge field $A$ in $3+1$ space-time dimensions is coupled to a complex scalar $\phi$ with potential $V(\phi) = - m^2 |\phi|^2 + \la |\phi|^4$, ($m^2 , \la > 0$) and Lagrangian
	\beq
	{\cal L} = \half ({\bf E}^2 - {\bf B}^2) + |(\pdr_\mu - i g A_\mu) \phi|^2 -  V(\phi).
	\eeq
The space of scalar vacua $\cal M$ (global minima of $V$) is a circle of radius $\eta = \sqrt{m^2/2\lambda}$. If U$(1)$ were a global symmetry we would have one angular Goldstone mode. A non-zero vacuum expectation value (vev) $\bra \phi \ket = \eta$ leads to complete spontaneous breaking of the symmetry group $G$. If $\phi = (\eta + \rho) e^{i \chi/\eta}$, we may gauge away $\chi$ and get a mass term $g^2 \eta^2 A^2$ for the photon (which has `eaten' the Goldstone mode), and a radial scalar mass term $m^2 \rho^2$ corresponding to the Higgs particle. In general \cite{non-Abelian-Higgs-Kibble}, $G$ breaks to a residual symmetry group $H$ whose generators annihilate the vacuum and $g^2 \eta^2 A^2$ is replaced by gauge boson mass terms $\half M_{ab} A_\mu^a A_\mu^b$. {\it We say that a spontaneously broken gauge theory corresponds to a rigid body, if vector boson masses and added-mass eigenvalues coincide.} In particular, the dimension of $G$ must equal that of the flow domain. We begin with some examples of SSB patterns and associated rigid bodies. In these examples, the space of scalar vacua ${\cal M}$ is the quotient $G/H$. They reveal a relation between symmetries of $G/H$ and of a corresponding {\it ideal} rigid body. By an ideal rigid body, we mean one with maximal symmetry group among those with identical added-mass eigenvalues: for example, a round sphere of appropriate radius, instead of a cube.

\begin{enumerate}

\item Consider an SO$(3)$ gauge theory minimally coupled to a triplet of real scalars interacting via the above potential $V$. ${\cal M}$ is a $2$-sphere of radius $\eta$ resulting in two Goldstone modes. They are eaten by $2$ of the $3$ gauge bosons leaving one massless photon. The mass-squared matrix $M$ is $2 g^2 \eta^2 \: {\rm diag}(1,1,0)$. $G=$SO$(3)$ breaks to $H = $SO$(2)$. The corresponding rigid body moves in fluid filling  three dimensional Euclidean space, since $\dim G = 3$. The rigid body must have one zero and two equal added-mass eigenvalues to correspond to the mass matrix $M$. An ideal rigid body that does the job is a hollow cylindrical shell, say $S^1 \times [-1,1]$. Such a shell has no added-mass when accelerated along its axis. Due to its circular cross section, the added-masses are equal and non-zero for acceleration in all directions normal to the axis.

\item Similarly, an SO$(n)$ gauge theory coupled to $n$-component real scalars spontaneously breaks to $H=$SO$(n-1)$. The vacuum manifold ${\cal M}$ is a sphere S$^{n-1}$ of radius $\eta$. We get $n-1$ vector bosons of mass $\sqrt{2} g \eta$ and $n_\gamma = {(n-1)(n-2)/2}$ massless photons. A corresponding ideal rigid body moving through fluid filling $\mathbb{R}^{\half (n^2 - n)}$ is the product $S^{n-2} \times B^{n_\g}$, generalizing the cylindrical shell $S^1 \times B^1$ when $n=3$. Here $B^{n_\g}$ is a unit ball $|{\bf x}| \leq 1$  for ${\bf x} \in \mathbb{R}^{n_\g}$. This ideal rigid body has equal non-zero added-masses when accelerated along the first $n-1$ directions and no added-mass in the remaining $n_\gamma$ {\em flat} directions. We call $S^{n-2}$ its curved factor and the unit ball $B^{n_\g}$ its flat factor. $B^1$ is the unit interval while $B^2$ is the unit disk, etc. It is easily seen that for $n = 3$ and $4$, acceleration along the direction of the interval or in the plane of the disk displaces no fluid, the same holds for $n \geq 5$.

\item For SU$(2)$ gauge fields coupled to a complex scalar doublet with the same potential $V$, ${\cal M}$ is a $3$-sphere of radius $\eta$. All $3$ gauge bosons are equally massive. The mass-squared matrix is $M_{ab} = (g^2 \eta^2/2) \: \delta_{ab}$ and SU$(2)$ breaks completely. A corresponding ideal rigid body is a $2$-sphere of radius $a= (3 g \eta/2 \pi \sqrt{2} \rho)^{1/3}$ moving through a fluid in three dimensions. The same group with scalars in other representations could lead to different SSB patterns and rigid bodies. With adjoint scalars, SU$(2) \to$ U$(1)$ with two equally massive vectors, corresponding to a hollow cylindrical shell moving in 3d.

\item In unbroken gauge theories, all gauge bosons remain massless. Such a theory with $d$-dimensional gauge group, corresponds to a point particle (or one of codimension more than one) moving through $\mathbb{R}^d$, which has no added mass. For instance, SU$(2)$ coupled to a complex scalar triplet in the potential $V = m^2 |\phi|^2 + \lambda |\phi|^4$ with $m^2 > 0$, remains unbroken and corresponds to a point particle moving through $\mathbb{R}^3$.

\item SU$(3)$ with fundamental scalars breaks to SU$(2)$ and ${\cal M} = S^5$. There are $3$ massless photons, $4$ vector bosons of mass $g \eta/\sqrt{2}$ and a heavier singlet of mass $\sqrt{2} g \eta/\sqrt{3}$. The corresponding ideal rigid body moves in $\mathbb{R}^8$. Its curved factor is a 4d ellipsoid $\sum_{i=1}^4 \frac{x_i^2}{a^2} + \frac{x_5^2}{b^2} = 1$ with $b < a$. The unit ball $B^3$ is its flat factor, which gives rise to three vanishing added-mass eigenvalues $\mu_6=\mu_7=\mu_8 =0$. Acceleration along the first five coordinates $x_1, \ldots, x_4, x_5$ leads to added-mass eigenvalues $\mu_1 = \ldots = \mu_4 < \mu_5$ since the semi-axes satisfy $a > b$ (higher added-mass when larger cross-section presented).

\item A U(1) gauge theory coupled to a complex scalar with charge $g n$ ($\phi \to e^{i n \tht(x)} \phi$) breaks completely in the above potential $V$, leaving one vector boson with mass $\sqrt{2} g n \eta$. The corresponding rigid body can be regarded as an arc of a circle moving through fluid flowing around the circumference, as in section \ref{s-added-mass-effect}.

\item Another illustrative class of theories have $G = $U$(1)^d$ with couplings $g_1, \ldots, g_d$ and $p$ complex scalars in a reducible representation ($p \leq d$ ensures all Goldstone modes are eaten). We assume the scalar $\phi_j$ has charge $q_{jk}$ under the $k^{\rm th}$ U$(1)$ factor and transforms as $\phi_j \to e^{i q_{jk} \tht_k(x)} \phi_j$. They are subject to the potential $\sum_{i=1}^{p} \left(- m_i^2 |\phi_i|^2 + \la_i |\phi_i|^4\right)$. If $\eta_i = (m_i^2/2\lambda_i)^{1/2}$, the vacuum manifold is a $p$-torus, the product of circles of radii $\eta_i$: ${\cal M} = S^1_{\eta_1}\times\ldots\times S^1_{\eta_p}$. There are $p$ Goldstone modes and the mass-squared matrix $M_{ab} = 2 \sum_{j=1}^p \eta_j^2 q_{ja} g_a q_{jb} g_b$ is a sum of $p$ rank-one matrices and generically has $d - p$ zero eigenvalues; $G=U(1)^d$ breaks to $U(1)^{d-p}$. A corresponding ideal rigid body moving in $\mathbb{R}^d$ generalizes the cylinder with elliptical cross-section. It is a product of a (curved) ellipsoid with a (flat) unit ball: $\{ \sum_{i=1}^{p}x_i^2/a^2_i = 1\} \times B^{d-p}$. For pairwise unequal $a_i$, it has distinct non-zero added-mass eigenvalues when accelerated along $x_1, \ldots, x_p$ and none along its $d-p$ flat directions. E.g., a U$(1)^3$ theory with a complex doublet in the above reducible representation breaks to U$(1)$. The corresponding rigid body is a cylinder with elliptical cross-section moving in $\mathbb{R}^3$. On the other hand, with $3$-component complex scalars, U$(1)^3$ completely breaks leaving $3$ massive vector bosons with generically distinct masses. A corresponding ideal rigid body is an ellipsoid moving through fluid filling $\mathbb{R}^3$.

\item It is interesting to identify the rigid body corresponding to electroweak symmetry breaking. Here $G =$ SU$(2)_L \times$ U$(1)_Y$ and $H = $U$(1)_{\rm Q}$ with a massless photon and $m_{W^+} = m_{W^-} < m_Z$. The corresponding rigid body must move through fluid filling $\mathbb{R}^4$, and have principal added-masses $\mu_1 = \mu_2 < \mu_3$, $\mu_4 = 0$. An ideal rigid body generalizes a hollow cylinder. It is the 3d hypersurface $\{\sum_{i=1}^3 x^2_i/a^2_i = 1\} \times [-1,1]$ with $a_1 = a_2 > a_3 > 0$, embedded in $\mathbb{R}^4$. It has an ellipsoid of revolution as cross-section. When accelerated along $x_4$, it displaces no fluid, but has equal added-masses when accelerated along $x_1$ and $x_2$.

\end{enumerate}

More generally, we may associate an ideal rigid body to any pattern $G \to H$ of SSB, through its vector boson mass-squared matrix $M_{ab}$. $M_{ab}$ can always be block diagonalized into a $p \times p$ non-degenerate block (whose eigenvalues $m_1^2, \ldots, m_p^2$ are the squares of the masses of the massive vector bosons) and a $(d-p)\times(d-p)$ zero matrix corresponding to massless photons, where $\dim G = d$ and $\dim H = d-p$. A corresponding ideal rigid body is a product of curved and flat factors. To the non-degenerate part of $M_{ab}$ we associate a `curved' $p-1$ dimensional ellipsoid $\frac{x_1^2}{a_1^2} + \ldots + \frac{x_p^2}{a_p^2} = 1$. The semi-axis lengths $a_i$ are fixed by the vector boson masses. The `flat' factor of the body can be taken as a $(d-p)$ dimensional unit ball $B^{d-p}: \left\{ x_{p+1}, \ldots , x_d \: | \: x_{p+1}^2 + \ldots + x_{d}^2 \leq 1 \right\}$. For $p=d-1$ it is an interval and for $p = d-2$ it is a unit disk etc. Motion along the flat directions $x_{p+1} \ldots x_d$ does not displace fluid, leading to $d-p$ zero added mass eigenvalues while acceleration in the first $p$ directions leads to $p$ non-zero added mass eigenvalues. If the vector boson masses are ordered as $0 < m_1 = m_2 = \ldots = m_{p_1} < m_{p_1 +1} = \ldots = m_{p_1 + p_2} < \ldots < m_{p - p_r + 1} = \ldots = m_p$, then the corresponding semi-axes of the ellipsoid satisfy $0 > a_1 = a_2 = \ldots = a_{p_1} > a_{p_1 +1} = \ldots = a_{p_1 + p_2} > \ldots > a_{p - p_r + 1} = \ldots = a_p$ since the added mass grows with cross-sectional area presented. Here we have allowed for degeneracies among the masses, so that there are $r$ distinct non-zero masses with degeneracies $p_1, \ldots, p_r$ and $p = p_1 + \ldots + p_r$. To find an explicit formula for the semi-axes $a_i$ in terms of the vector boson masses and fluid density $\rho$, we would need to solve the potential flow equations around this rigid body.

%------------------------
\subsection{Symmetries of $G/H$ and of Rigid Body}
\label{s-symm-GmodH-and-rigid-body}
%------------------------

In all these examples, the ideal rigid body corresponding to a given pattern of symmetry breaking is a product of curved and flat factors, with added-mass for acceleration along the former. The flat factor could be taken as an interval/disk/ball of dimension $\dim H$. The vacuum manifold ${\cal M} = G/H$ could be endowed with a non-degenerate metric determined by the vector boson mass-squared matrix $M_{ab}$, since in all these examples, the number of Goldstone modes $p = \dim {\cal M}$ is equal to the number of massive vector bosons. $M_{ab}$ is in general degenerate, but may be block diagonalized into a non-degenerate $p \times p$ block and a zero matrix (corresponding to residual symmetries in $\underline{H}$). The non-degenerate part defines a metric $g$ on the quotient $G/H$. $G/H$ is a homogeneous space, so consider any point $m$ and define its `group of symmetries' $\cal G$ as the subgroup of $O(p)$ that fixes the metric at $m$, i.e, $R^t R = I, R^t g R = g$. So $\cal G$ are orthogonal symmetries of the metric in the tangent space $T_m(G/H)$. By homogeneity, $\cal G$ is independent of the chosen point $m$. Then $\cal G$ coincides with the group of rotation and reflection symmetries of the curved factor of the corresponding ideal rigid body. So the group $\cal G$ consists of symmetries of both the vector boson `mass metric' and the Euclidean metric in the flow domain inhabited by the rigid body. Let us illustrate this equality of symmetry groups in the above examples, the results are summarized in Table 1. To identify the group of symmetries in each case, we go to a basis in which the mass metric $g$ at a given point $m$ on $G/H$ is diagonal $g = {\rm diag}(\la_1, \ldots, \la_p)$. The eigenvalues are ordered as 
	\beq
	0 < \la_1 = \ldots = \la_{p_1} < \la_{p_1 + 1} = \ldots = \la_{p_1+p_2} < \ldots < \la_{p-p_{r}+1} = \ldots = \la_{p}
	\eeq
with $p = p_1 + \ldots + p_r$. Then one checks that the subgroup of O$(p)$ that commutes with $g$ is O$(p_1) \times$ O$(p_2) \times \cdots \times$ O$(p_r)$, with O$(1) = \mathbb{Z}_2$.

\begin{enumerate}

\item[(A)] If $G = $SU$(2)$, then ${\cal M} = S^3$ with round metric (all three eigenvalues equal), and the group of symmetries ${\cal G} = $O$(3)$ is maximal. The corresponding ideal rigid body $S^2$ has the same isometry group O$(3)$. 

\item[(B)] If $G= $SO$(p+1)$, ${\cal M} = $S$^{p}$ is round and ${\cal G} =$ O$(p)$, coinciding with the isometry group of the curved factor ${\rm S}^{p-1}$ of the corresponding rigid body.

\item[(C)] Suppose $G = $ U$(1)^d$ with $p$ scalars as above. Then ${\cal M}$ is a $p$-torus, generically with circles of distinct radii $\eta_i$. The symmetry group at $m \in {\cal M}$ is generated by reflections about $m$ along the $p$ circumferences, so ${\cal G} = (\mathbb{Z}_2)^p$. $\cal G$ coincides with the symmetry group (generated by $x_i \to - x_i$) of the ellipsoid factor $\{ \sum_{i=1}^p x_i^2/a_i^2 = 1 \}$ in the corresponding ideal rigid body. If two radii $\eta_1, \eta_2$ coincide, then ${\cal G} = $O$(2) \times (\mathbb{Z}_2)^{p-2}$ which agrees with the symmetries of an ellipsoid of revolution $(x_1^2/a_1^2 + x_2^2/a_1^2 + x_3^2/a_3^2 + \cdots + x_p^2/a_p^2 = 1)$, which is the curved factor of the corresponding ideal rigid body. 

\item[(D)] If $G =$ SU$(2) \times $U$(1)$ of the electroweak standard model, then ${\cal M} = $S$^3$. The metric is not round, as $m_W \ne m_Z$. ${\cal G} = $O$(2) \times \mathbb{Z}_2$, coinciding with the symmetry group of the curved factor of the corresponding rigid body.

\item[(E)] If $G = $SU$(3)$ then ${\cal M} = $S$^5$ with non-round metric and the symmetry group on either side is O$(4) \times \mathbb{Z}_2$ corresponding to the five non-zero added-masses $\mu_1 = \ldots = \mu_4 < \mu_5$.

\end{enumerate}

\begin{table}
\begin{center}

\begin{tabular}{| c | c | c | c | c | c | c |} \hline
\textbf{Gauge group $G$} & \textbf{Repn.} & ${\cal M} = G/H$ & $H$ & \textbf{Fluid} & \textbf{Ideal Rigid Body} & $\cal G$
\\ \hline

U(1) & 1d cx & $S_\eta^1$ & $\{1\}$ & $S^1$ & Arc $[\tht_1,\tht_2]$ & $Z_2$
\\ \hline

U$(1)^2$ & 2d cx & $S_{\eta_1}^1 \times S_{\eta_2}^1$ & $\{1\}$ & R$^2$ & Elliptical disk & $Z_2 \times Z_2$
\\ \hline

U$(1)^3$ & 2d cx & $S_{\eta_1}^1 \times S_{\eta_2}^1$ & U$(1)$ & R$^3$ & \begin{tabular}[c]{@{}c@{}} Hollow\\ elliptical cylinder\end{tabular} & $Z_2 \times Z_2$
\\ \hline

U$(1)^3$ & 3d cx & \begin{tabular}[c]{@{}c@{}}$S_{\eta_1}^1 \times S_{\eta_2}^1$ \\ $\times S_{\eta_3}^1$\end{tabular} & $\{1\}$ & R$^3$ & Ellipsoid & $(Z_2)^3$
\\ \hline

SU(2) & 2d cx & $S_{\eta}^3$ & $\{1\}$ & R$^3$ & Sphere & $O(3)$
\\ \hline

SO(3) & 3d rl & $S_{\eta}^2$ & SO(2) & R$^3$ & \begin{tabular}[c]{@{}c@{}} Hollow\\ circular cylinder\end{tabular} & $O(2)$
\\ \hline

SU$(2)_L \times$ U$(1)_Y$ & 2d cx & $S_{\eta}^3$ & U$(1)_{\rm Q}$ & R$^4$ & Ellipsoid $\times$ $[a,b]$ & \begin{tabular}[c]{@{}c@{}}$O(2) \times Z_2$ \end{tabular}
\\ \hline

SU(3) & 3d cx & $S_{\eta}^5$ & SU(2) & R$^8$ & \begin{tabular}[c]{@{}c@{}} 4d ellipsoid\\ $\sum_{i=1}^4\frac{x_i^2}{a^2} + \frac{x_5^2}{b^2}$ \end{tabular} & $O(4) \times Z_2$
\\ \hline

Any $G$ of dim $d$ & $d$-dim cx & $\phi = 0$ & $G$ & R$^d$ & Point particle & $O(d)$
\\ \hline

\end{tabular}

\label{t-eg-ssb-pattern-body}

\caption{Patterns of spontaneous symmetry breaking and corresponding rigid bodies for various gauge groups $G$ and scalar field representations (real - rl or complex - cx). The vacuum manifold $\cal M$, residual symmetry group $H$, fluid flow domain, ideal rigid body and group of symmetries $\cal G$ of the curved factor of the body are listed. The Higgs potential in the case of a point particle is $V = m^2 |\phi|^2 + \lambda |\phi|^4$, while in all other cases $V = -m^2 |\phi|^2 + \lambda |\phi|^4$ as in the text. These results hold for generic values of charges $q_{ij}$, vev $\eta_i$ and gauge couplings. $S^n_\eta$ denotes an $n$-sphere of radius $\eta$.}

\end{center}

\end{table}

% \vspace{-.5cm}

%================================================
\section{The Higgs Added-Mass Correspondence}
\label{s-Dictionary}
%================================================

% \vspace{-.4cm}

\begin{table}[h]
\begin{center}
\begin{tabular}{| p{5cm} | p{5cm} |} \hline
{\textbf{Added-Mass Effect}} & {\textbf{Higgs Mechanism}} \\ \hline
 
Rigid body & Gauge bosons \\ \hline
Fluid & Scalar field \\ \hline
Space occupied by fluid & Gauge Lie algebra \\ \hline

Dimension of container & $\dim G$ \\ \hline

Added-mass tensor $\mu_{ij}$ & Mass matrix $M_{ab}$ \\ \hline

Added-mass eigenvalues & Vector boson masses \\ \hline

Acceleration along flat face & Massless photon \\ \hline

Number of flat directions & $\dim H$ \\ \hline

Sphere moving in 3d & SU$(2) \to \{ 1 \}$, doublet \\ \hline
Hollow cylinder in 3d & SO$(3) \to$ SO$(2)$, triplet \\ \hline
Broken pressure symmetry & Broken gauge symmetry \\ \hline
Fluid density $\rho$ & vev $\bra \phi \ket$ of Higgs scalar \\ \hline
$F_i - ma_i \; = \; \mu_{ij} \; a_j $ & $-j^{\nu} + \partial_{\mu}F^{\mu\nu} = g^2 \bra \phi \ket^2 A^{\nu}$ \\ \hline
boundary condition on body surface & Gauge-scalar coupling \\ \hline
Long wavelength fluid mode & Higgs particle \\ \hline

% Acceleration vector $a_i$ & Couplings $g_i$ in U$(1)^d$ \\  \hline % could omit

% Dimensions of body & Charges of scalar in U$(1)^d$ \\ \hline % could omit

Symmetries of curved body & Symmetries of $T_m ({G/H})$ \\ \hline

\end{tabular}

\label{t-table-HAM}

\caption{The Higgs added-mass correspondence.}

\end{center}
\end{table}

We now mention some striking analogies between the added-mass effect and Higgs mechanism. They are summarized in Table 2. The rigid body plays the role of gauge bosons, both can gain mass. The fluid plays the role of the scalar field. When the body is accelerated, some energy goes into the flow. Figuratively, the body carries fluid, adding to its mass. Similarly, gauge bosons gain mass by carrying Goldstone modes. The analogy relates the space of fluid flow, to the Lie algebra $\underline{G}$ (the location of the body provides an origin for the flow domain and it is the space of directions in which the body can move that corresponds to the gauge Lie algebra). The dimension $d$ of the fluid container $\mathbb{R}^d$ equals $\dim \underline{G}$. The added-mass tensor $\mu_{ij}$ and the vector boson mass-squared matrix $M_{ab}$ are both $d \times d$ matrices. A direction of acceleration relative to the body is equivalent to a direction in $\underline{G}$. Zero modes of $\mu_{ij}$ are directions in which the acceleration reaction force vanishes. These are like directions of residual symmetry in the Lie algebra $\underline{H}$. A thin disk accelerated along its surface gains no added-mass when moving in a 3d fluid, just as we have a massless photon along an unbroken symmetry generator of $G$. In general, the number of flat directions of the body is equal to the number of massless vectors.

We say that a particular spontaneous symmetry breaking pattern {\it corresponds} to a particular rigid body if the vector boson masses coincide with the added-mass eigenvalues. The latter do not, generally, determine the body. A sphere and cube of appropriate sizes have identical added-mass eigenvalues, just as appropriate SU$(2)$ and U$(1)^3$ gauge theories share vector boson mass spectra. So the correspondence, at this level, relates a class of classical gauge theories to a family of rigid bodies. Among these rigid bodies there are `ideal' ones, with maximal symmetry group. The identification of $\underline{G}$ with the space of fluid flow related symmetries of the `mass' metric at any point of $G/H$ to those of the curved factor of the corresponding ideal rigid body (see \S \ref{s-symm-GmodH-and-rigid-body}).

Consider a bounded rigid body that moves at constant velocity through an infinite, inviscid, incompressible, irrotational potential flow without the formation of vortex sheets, wakes or cavities. It feels no added-mass (this is part of d'Alembert's `paradox' \cite{Batchelor}). However, it is associated with a `benign' flow not requiring energy input. For example, the flow field around a uniformly moving sphere of radius $a$, instantaneously centered at $U t \hat z$ is
	\beq
	\bfv(\bfr,t) = \frac{a^3 U}{2r'(t)^3} \left[2\cos \tht'(t) \,\hat \bfr'(t) + \sin \tht'(t) \, \hat\tht'(t) \right]
	\label{e:vel-around-sphere}
	\eeq
where $\bfr' = \bfr - U t \hat z$ is the position vector of the observation point relative to the center of the sphere. So a body moving steadily is not coupled to the fluid through energy exchange. Similarly, if the scalar vacuum expectation value $\bra \phi \ket$ is non-zero but the gauge coupling $g$ is zero, then we have spontaneous symmetry breaking and Goldstone modes, but massless gauge bosons. The Goldstone modes are analogous to the above benign flow.

Is there a broken symmetry in the added-mass effect? When a sphere moves uniformly, from (\ref{e:vel-around-sphere}) and Bernoulli's equation (\ref{e-bernoulli-eqn}), the pressure distributions on the front and rear hemispheres are identical. This front-back symmetry is broken upon accelerating the sphere. It is a discrete analog of the broken gauge symmetry. Moreover, spontaneous symmetry breaking is caused by a non-zero vacuum expectation value $|\bra \phi \ket| = \eta$. The density $\rho$ is its counterpart. Both occur as pre-factors in mass matrices ($\mu^{\rm sphere}_{ij} \propto \rho a^3 \del_{ij}$, $M^{{\rm SU(2)}}_{ab} \propto \eta^2 g^2 \del_{ab}$) and are exclusively properties of fluid and scalar field (i.e., not having to do with the rigid body or gauge fields).

Our analogy extends to the dynamical equations of the body ($F_i - ma_i = \mu_{ij} a_j $) and massive vector boson ($-j^{\nu} + \partial_{\mu}F^{\mu\nu} = g^2 \eta^2 A^{\nu}$). The added-mass $\mu_{ij}a_j$ is like the Proca mass. The external force $F_i$ and current $j^\nu$ are both sources in otherwise homogeneous equations. $\partial_{\mu}F^{\mu\nu} = 0$ is the analog of $m a_i = 0$: free propagation of electromagnetic waves is like uniform motion of a rigid body. The impenetrable body-fluid boundary condition is analogous to gauge-scalar minimal coupling. Other boundary conditions would correspond to non-minimally coupled scalars.

From the spontaneously broken U$(1)^d$ models of \S \ref{ssb-patters-rigid-bodies}, we obtain further analogies. There are $3$ ways to prevent spontaneous gauge symmetry breaking: (a) set the gauge couplings $g_i$ to zero, (b) make the scalars uncharged ($n_{ij} \to 0$) under U$(1)^d$ and (c) let the scalar vacuum expectation value $\eta \to 0$. Similarly, there are $3$ ways to make the added force/mass vanish: (a) set the acceleration components $a_i = 0$, (b) shrink the body to a point and (c) let $\rho \to 0$.

% \vspace{-.5cm}
%==================================================
\section{Discussion}
\label{Discussion}
%==================================================
% \vspace{-.4cm}

In this paper, we have proposed a new physical correspondence between the Higgs mechanism in particle physics and the added-mass effect in fluid mechanics. While plasmas and superconductors illustrate the Abelian Higgs model, the Higgs added-mass correspondence provides a non-dissipative hydrodynamic analogy for the fully non-Abelian Higgs mechanism. It encodes a pattern of gauge symmetry breaking in the shape of a rigid body accelerated through fluid. A dictionary relates symmetries and various quantities on either side. By identifying the gauge Lie algebra with the space of fluid flow, and relating added-mass eigenvalues to vector boson masses, we are able to specify when a particular pattern of spontaneous symmetry breaking {\it corresponds} to a particular rigid body accelerated through a fluid. Besides possible refinements and generalizations (to compressible [see \S \ref{app-added-mass-compress-potn-flow}] and rotational flows or inclusion of fermion masses), the new viewpoint raises several interesting questions and directions for further research in both fluid mechanics and particle physics (1) The Higgs is the lightest scalar particle. We conjecture that the fluid analog is a characteristic fluid mode around an accelerating body, with wave length comparable to the size of the body (rather than the container). There may be several such modes, which could suggest heavier scalar particles. (2) Understanding such modes requires extension of the added mass formalism to flows other than those usually studied in marine hydrodynamics (incompressible potential flow). This would allow for waves around the body, that could play the role of the Higgs particle. Perhaps the simplest such flows are compressible potential flow, incompressible flows with vorticity (even in two dimensions) and surface gravity waves in incompressible flow around an accelerated body. Moreover, density fluctuations in compressible flow around a rigid body should be analogous to quantum fluctuations around the scalar vacuum expectation value. Thus the HAM correspondence gives a new viewpoint and impetus to develop techniques to study the added mass effect in flows other than those studied so far. (3) We identified a discrete broken symmetry in the added-mass effect. Is there a continuous one, perhaps having to do with Galilean invariance? (4) The fluid flow affects the rotational inertia of a rigid body, giving it an added inertia tensor. Is there a particle physics analog consistent with the quantization of angular momentum? For instance, could motion through the scalar medium modify the magnetic moments of particles? (5) The HAM correspondence relates rigid body motion through $d$-dimensional flows (see \S \ref{app-d-dim-addedmass}) to SSB of gauge theories with $d$-dimensional gauge group. Given the importance and simplifications in the 't Hooft limit of multi-color gauge models, one wonders whether there are aspects of these fluid flows that simplify as $d \to \infty$. Could a suitable $d \to \infty$ limit provide a starting point for an approximation method for studying 3d flows? (6) How is the added-mass of a composite body related to the added masses of its constituents? Correspondingly, can one compute a small correction to the mass of a hadron, from Higgs interactions among a system of quarks [beyond the Higgs contribution to individual current quark masses]? This would be a small `Higgs force' correction to the mass of the proton in addition to the main contributions from strong and electromagnetic forces.

\section{Acknowledgements} We thank A. Thyagaraja, V. V. Sreedhar, K. G. Arun, N. D. Hari Dass, T. R. Ramadas, A. Laddha, G. Date, G. Rajasekaran and R. Nityananda for useful discussions, and anonymous referees for suggesting improvements to the manuscript. This work was supported in part by grants from the Infosys Foundation and the Dept. of Science \& Technology, Govt. of India.

%-----------------------
\appendix

\section{Added mass effect in $d \geq 3$ dimensions}
\label{app-d-dim-addedmass}
%-----------------------

The HAM correspondence relates spontaneous breaking of a $d$-dimensional gauge group $G$ to the added mass effect in $d$-dimensional fluids. Since there is no restriction on the dimension of $G$, our correspondence requires an extension of the standard added mass effect \cite{Batchelor,Kambe} to flows in $d \geq 4$, which we give here. Consider incompressible potential flow in $\mathbb{R}^d$ around a simply connected rigid body moving with velocity $\bfU(t)$. We assume that the body executes purely translational motion and that $\bfv \to 0$ asymptotically. The velocity potential satisfies the Laplace equation $\grad^2 \phi = 0$ subject to impenetrable boundary conditions on the body surface: $\bfn \cdot \grad \phi = \bfn \cdot \bfU$. With the origin located inside the body, $\phi$ admits a multipole expansion in terms of Green's function for the laplacian $\grad^2 g(r) = \del^d(\bfr)$, $g(r) = -\frac{\Gamma(d/2)}{2\pi^{d/2}(d-2)} \frac{1}{r^{d-2}}$ and its derivatives:
	\beq
	\phi(\bfr) = c/r^{d-2} + c_i \pdr_i (1/r^{d-2}) + c_{ij} \partial_i \partial_j ( 1/r^{d-2}) + \ldots \label{e-multipole-expansion}
	\eeq
As in the Cauchy contour integral formula, the multipole tensor coefficients (which are linear in $\bfU$) may be expressed as integrals of $\phi$ and its derivatives over the body surface $A$,
	\beqs
 	c &=& \frac{\Gamma(\frac{d}{2})}{2\pi^{d/2}(d-2)} \oint_{A} {\bf n} \cdot \grad \phi(\bfr) \, dA, \quad 
 	c_i = \frac{\Gamma(\frac{d}{2})}{2\pi^{d/2}(d-2)} \, \oint_{A} \left[ (\bfn \cdot \grad \phi) r_i - \phi n_i \right]\,dA, \cr
	c_{ij} &=& \frac{\Gamma(\frac{d}{2})}{2\pi^{d/2}(d-2)} \oint_{A} \left[ (\bfn \cdot \grad \phi) r_i r_j - \phi (n_i r_j + n_j r_i) \right]\, dA, \quad \ldots 
	\label{e-multipole-moments}\label{e-multipole-coefficients}
	\eeqs
For incompressible flow without sources, the monopole coefficient $c \equiv 0$. As in the $3$d case, the impenetrable boundary condition constrains $\phi$ to be linear in $\bfU$, which allows us to write it as $\phi = {\bf \Phi}\cdot\bfU$. The potential vector field ${\bf \Phi}(\bfr, t) = {\bf \Phi}(\bfr - \bfr_0(t))$ is independent of $\bf U$. $\bfr_0(t)$ is a convenient reference point fixed in the body. As in \S \ref{s-added-mass-effect}, we use Bernoulli's equation (\ref{e-bernoulli-eqn}) to write the pressure-force on the body surface $A$ in terms of $\phi$, and use the factorization $\phi = {\Phi}\cdot\bfU$ to write the force as the sum of an acceleration reaction $\bf G$ and a non-acceleration force ${\bf G}'$, as in (\ref{e-G-G-prime}). From the multipole expansion $\phi \sim 1/r^{d-1}$ and it follows that $\bfG'$ vanishes when the flow domain is all of $\mathbb{R}^d$. Thus we get the same formula (as in 3d) for the added mass tensor $\mu_{ij}$ from the acceleration-reaction force:
	\beq
	G_i \;=\; \rho\, \dot{U_j} \oint_{A} \Phi_j\, n_i \, dA \;\equiv\; -\mu_{ij}\,\dot{U_j} \quad \Rightarrow \quad \mu_{ij} = -\rho \oint_{A} \Phi_j\, n_i \, dA.
	\eeq
Despite appearances, $\mu_{ij}$ only depends on the dipole term in $\phi$. The linearity of the boundary condition in $\bf U$ implies that $c_i = d_{ij} U_j$ is linear in $\bf U$. The constant source doublet/dipole tensor $d_{ij}$ depends only on the shape of the body. Using eqn. (\ref{e-multipole-coefficients}) for $c_i$ and the boundary condition on the surface, we obtain
	\beqs
	c_i &=& d_{ij}U_j = \frac{\Gamma(\frac{d}{2})}{2(d-2)\pi^{d/2}} \oint_{A} \left[ (\bfn\cdot\bfU)r_i - \phi n_i \right]\, dA \cr 
	&=& \frac{\Gamma(\frac{d}{2})}{2(d-2)\pi^{d/2}} U_j \left[ \int_{{\rm body}} \partial_j r_i\, dV - \oint_{A} \Phi_j n_i\, dA \right] 
	= \frac{\Gamma(\frac{d}{2})}{2(d-2)\pi^{d/2}} \left[ V_{\rm body} \del_{ij} + \frac{\mu_{ij}}{\rho} \right]U_j.
	\eeqs
Since this is valid for any velocity $\bf U$ we arrive at a relation between $\mu_{ij}$ and the dipole tensor
	\beq
	\mu_{ij} = \rho \left[ \frac{2(d-2) \, \pi^{d/2}}{\Gamma(d/2)} \: d_{ij} - V_{\rm body} \del_{ij} \right].
	\eeq
This expression for $\mu_{ij}$ shows that it only depends on the dipole part of $\phi$. It does not involve integrals and gives a simple way of computing $\mu_{ij}$ once the dipole term in $\phi$ is known. Let us illustrate this with the example of a $(d-1)$-dimensional sphere $S^{d-1}_R$ of radius $a$, moving through fluid in $\mathbb{R}^d$. A moving sphere instantaneously centered at the origin induces a dipole flow field with potential $ \phi = c_i \pdr_i r^{2-d} = -(d-2) r^{-d} \:  {\bf c} \cdot {\bf r}$. The multipole tensors $c_{ij}, c_{ijk}, \ldots$ are {\rm constant} tensors of rank $> 1$, linear in $\bfU$. Spherical symmetry of the body denies us any other vector/tensor from which to construct them, so they must vanish. The dipole coefficient $\bf c$ may be self-consistently determined by inserting this formula for $\phi$ in (\ref{e-multipole-moments}). One obtains
	\beq
	c_i = \frac{a^d}{(d-1)(d-2)} U_i \qquad
	\text{or} \qquad
	d_{ij} = \frac{a^d}{(d-1)(d-2)} \, \del_{ij}.
	\label{e-sphere-dipole-moment} 
	\eeq
Hence, the added mass tensor for a $(d-1)$-sphere of radius $a$ moving in $\mathbb{R}^d$ is
	\beq
	\mu_{ij}^{\rm sphere} \;=\; \rho \frac{2 \pi^{d/2} a^d}{d(d-1) \Gamma\left( \frac{d}{2}\right)} \del_{ij}
	= \frac{\text{(Mass of fluid displaced)}}{(d-1)} \del_{ij}.
	\eeq
This reduces to the well-known results for planar or 3d flow around a disk or $2$-sphere. In \S \ref{Discussion}, we speculate on the possible relevance of a suitable $d \to \infty$ limit.

%-----------------------
\section{Added mass effect for compressible potential flow}
\label{app-added-mass-compress-potn-flow}
%-----------------------

Treatments of the added mass effect assume for simplicity that the flow is inviscid, incompressible and irrotational. However, physically, it is clear that the effect is present even in compressible or rotational flow. Indeed, according to our correspondence, density fluctuations around incompressible flow should correspond to quantum fluctuations around a constant vev for the scalar field. Moreover, to look for a fluid analogue of the Higgs particle, i.e., a `Higgs wave' around an accelerated rigid body, we need a generalization of the added mass effect to compressible flow. As is well known, the resulting flows can be very complicated. Here we take a small step by formulating the added mass effect for compressible potential flow around a rigid body executing purely translational motion at velocity $\bfU(t)$. We assume the flow is isentropic so that $\grad p/\rho = \grad h$ where $h$ is specific enthalpy. Euler's equation $\dd{\bfv}{t} + \bfv \cdot \grad \bfv = - \grad h$ then implies an unsteady Bernoulli equation for the velocity potential $\phi$,
	\beq
	\pdr_t \phi + (1/2) \bfv^2 + h = {\rm constant}(t).
	\eeq
For concreteness, we consider adiabatic motion of an ideal gas so that $(p/p_0) = (\rho/\rho_0)^\gamma$ where $\gamma = c_p/c_v$ is the adiabatic index and $p_0,\rho_0$ are reference pressure and density. Then $h = [\gamma/(\gamma -1)] p/\rho$. Of course, $\phi$ and $\rho$ are to be determined by solving the Euler and continuity equations subject to initial and boundary conditions. To identify the added force on the body, it helps to regard the continuity equation and impenetrable boundary condition on the body, namely
	\beq
	(\grad \rho \cdot \grad + \rho \grad^2) \phi = - \pdr_t \rho \quad \text{and} \quad
	\hat n \cdot \grad \phi = \hat n \cdot \bfU,
	\eeq
as a system of inhomogeneous linear equations for $\phi$ given $\rho$ and $\bf U$. The rhs of this system is linear in $\bfU$ (and $\rho$), so formally, the solution of this equation can be expressed as $\phi = \bfU \cdot {\bf \Phi}(\bfr,t) + \psi(\bfr,t)$ where the potential vector field $\bf \Phi$ and the supplementary potential $\psi$ are $\bfU$-independent but depend on $\rho$. To see why this is true, discretize the system as a matrix equation $A(\rho) \phi = b$. The upper rows of the matrix $A$ encode the operator $\grad \rho \cdot \grad + \rho \grad^2$ while the lower rows encode $\hat n \cdot \grad$. The upper rows of the column vector $b$ represent $- \pdr_t \rho$ and the lower rows contain $\hat n \cdot \bfU$, so that we may write $b  = b_1(\rho) + b_2(\bfU)$ where $b_2$ is linear in $\bf U$. Inverting $A$ gives the desired decomposition.

With the aid of Bernoulli's equation, the force on the body $-\int_A p \, \hat n \, dA$ becomes
	\beq
	F_i = \left(\frac{\gamma -1}{\gamma} \right) \int_A \rho \left[ \pdr_t \phi + \half \bfv^2 - {\rm const}(t) \right] n_i dA.
	\eeq
Using our factorization $\phi = \bfU \cdot {\bf \Phi} + \psi$, the force on the body is the sum of an acceleration reaction force $G_i = - \mu_{ij} \dot U_j$ and a non-acceleration force ${\bf G}'$:
	\beq
	G_i = \frac{(\gamma-1)}{\gamma} \dot U_j \int_A \rho \Phi_j n_i \, dA \quad \text{and} \quad G'_i = \frac{\gamma-1}{\gamma} \int_A \rho \left[U_j \dot \Phi_j + \dot \psi + \frac{\bfv^2}{2} - {\rm const}(t) \right] n_i \, dA.
	\eeq
The added mass tensor $\mu_{ij} = -\frac{(\gamma-1)}{\gamma} \int_A \rho \Phi_j n_i \, dA$. To find $\mu_{ij}$ for a given body, we need to solve for $\rho$ and $\bfv$ using the equations of motion. Unlike for constant density, where $\mu_{ij}$ is constant, here it could change with time and location due to density variations arising from the acceleration of the body. Corrections to the added mass due to density fluctuations are analogous to corrections to the $W$ and $Z$ boson masses due to quantum fluctuations around a constant scalar vev. This interesting phenomenon will be further investigated elsewhere.

%============================================


\begin{thebibliography}{99}

%\cite{Aad:2012tfa}
\bibitem{ATLAS} 
  G.~Aad {\it et al.}  [ATLAS Collaboration],
``Observation of a new particle in the search for the Standard Model Higgs boson with the ATLAS detector at the LHC,''
  Phys.\ Lett.\ B {\bf 716}, 1 (2012)
%  [arXiv:1207.7214 [hep-ex]].
  %%CITATION = ARXIV:1207.7214;%%
  %2887 citations counted in INSPIRE as of 23 Jun 2014

%\cite{Chatrchyan:2012ufa}
\bibitem{CMS} 
  S.~Chatrchyan {\it et al.}  [CMS Collaboration],
``Observation of a new boson at a mass of 125 GeV with the CMS experiment at the LHC,''
  Phys.\ Lett.\ B {\bf 716}, 30 (2012)
%  [arXiv:1207.7235 [hep-ex]].
  %%CITATION = ARXIV:1207.7235;%%
  %2834 citations counted in INSPIRE as of 23 Jun 2014

\bibitem{Anderson} 
  P.~W.~Anderson,
  ``Plasmons, Gauge Invariance, and Mass,''
  Phys.\ Rev.\  {\bf 130}, 439 (1963).
  %%CITATION = PHRVA,130,439;%%
  %274 citations counted in INSPIRE as of 23 Jun 2014

\bibitem{EnglertBrout} 
  F.~Englert and R.~Brout,
  ``Broken Symmetry and the Mass of Gauge Vector Mesons,''
  Phys.\ Rev.\ Lett.\  {\bf 13}, 321 (1964).
  %%CITATION = PRLTA,13,321;%%
  %2504 citations counted in INSPIRE as of 23 Jun 2014
%\cite{Higgs:1964pj}

\bibitem{Higgs} 
  P.~W.~Higgs,
``Broken Symmetries and the Masses of Gauge Bosons,''
  Phys.\ Rev.\ Lett.\  {\bf 13}, 508 (1964).
  %%CITATION = PRLTA,13,508;%%
  %2749 citations counted in INSPIRE as of 20 Jun 2014

\bibitem{GuralnikHagenKibble}
G.~S.~Guralnik, C.~R.~Hagen and T.~W.~B.~Kibble,
  ``Global Conservation Laws and Massless Particles,''
  Phys.\ Rev.\ Lett.\  {\bf 13}, 585 (1964).
  %%CITATION = PRLTA,13,585;%%
  %1992 citations counted in INSPIRE as of 23 Jun 2014

\bibitem{Thyagaraja}  K. G. McClements and A. Thyagaraja,
``Understanding the Higgs Mechanism,''
Physics World, October 2012, 23

%  \bibitem{Green} G. Green, \emph{On the Vibration of Pendulums in Fluid Media}, Edin. Trans. 1833 [Papers, p.315]
%  \bibitem{Stokes}  G. G. Stokes, \emph{On the effect of the internal friction of fluids on the motion of pendulums}, Transactions of the Cambridge Philosophical Society, 9: 8-106, 1851

\bibitem{Lamb} H. Lamb, ``Hydrodynamics,'' Camb. Univ. Press, (1932).

\bibitem{Batchelor} G. K. Batchelor, ``An Introduction to Fluid Dynamics,'' Camb. Univ. Press, (1967).


\bibitem{non-Abelian-Higgs-Kibble}
T.~W.~B.~Kibble,
  ``Symmetry breaking in non-Abelian gauge theories,''
  Phys.\ Rev.\  {\bf 155}, 1554 (1967).
  %%CITATION = PHRVA,155,1554;%%
  %1043 citations counted in INSPIRE as of 23 Jun 2014

\bibitem{Kambe}
T.~Kambe, ``Elementary Fluid Mechanics,'' World Scientific Publishing, (2007).
  
\end{thebibliography}
\end{document}